\documentclass[12pt, preprint]{emulateapj}
\usepackage{multirow}
\usepackage{amsmath}
\renewcommand{\baselinestretch}{1.1}
\usepackage{natbib}

\newcommand       \Angstrom     {\,{\rm \AA}}

\newcommand       \g          {\,{\rm g}}
\newcommand       \cm          {\,{\rm cm}}

\newcommand       \Mpc          {\,{\rm Mpc}}

\newcommand       \um           {\,{\rm \mu m}}
\newcommand       \mum           {\,{\rm \mu m}}

\newcommand       \Rv           {R_V}
\newcommand       \Av           {A_V}

\newcommand       \HH           {{\rm H}}

\newcommand       \magni        {\,{\rm mag}}
\newcommand       \simali       {{\sim\,}}

\newcommand       \nH           {n_{\rm H}}
\newcommand       \NH           {N_{\rm H}}

\newcommand       \Bsil          {B_{\rm sil}}
\newcommand       \Bgra          {B_{\rm gra}}
\newcommand       \rhosil        {\rho_{\rm sil}}
\newcommand       \rhogra        {\rho_{\rm gra}}

\newcommand	  \xism         {\left[{\rm X/H}\right]_{\rm ISM}}
\newcommand	  \csun         {\left[{\rm C/H}\right]_{\odot}}

\newcommand	  \cism         {\left[{\rm C/H}\right]_{\rm ISM}}

\newcommand	  \feism        {\left[{\rm Fe/H}\right]_{\rm ISM}}
\newcommand	  \mgism        {\left[{\rm Mg/H}\right]_{\rm ISM}}
\newcommand	  \siism        {\left[{\rm Si/H}\right]_{\rm ISM}}
\newcommand	  \xdust        {\left[{\rm X/H}\right]_{\rm dust}}
\newcommand	  \cdust        {\left[{\rm C/H}\right]_{\rm dust}}

\newcommand	  \odust        {\left[{\rm O/H}\right]_{\rm dust}}
\newcommand	  \fedust       {\left[{\rm Fe/H}\right]_{\rm dust}}
\newcommand	  \mgdust       {\left[{\rm Mg/H}\right]_{\rm dust}}
\newcommand	  \sidust       {\left[{\rm Si/H}\right]_{\rm dust}}
\newcommand	  \xgas         {\left[{\rm X/H}\right]_{\rm gas}}
\newcommand	  \cgas        {\left[{\rm C/H}\right]_{\rm gas}}

\newcommand	  \mux         {\mu_{\rm X}}
\newcommand	  \muc         {\mu_{\rm C}}

\newcommand	  \musi        {\mu_{\rm Si}}
\newcommand	  \mufe        {\mu_{\rm Fe}}
\newcommand	  \mumg        {\mu_{\rm Mg}}

\newcommand	  \ppm	       {\,{\rm ppm}}
\newcommand	  \amin	       {a_{\rm min}}

\newcommand       \simlt        {\lesssim}

\countdef\decade=200
\advance\decade by \year
\countdef\hours=201
\advance\hours by \time
\divide\hours by 60
\countdef\mins=202
\advance\mins by \hours
\multiply\mins by 60
\multiply\hours by 100
\countdef\miltime=203
\advance\miltime by \hours
\advance\miltime by \time
\advance\miltime by -\mins
\def\today{\number\decade.\number\month.\number\day.\number\miltime}
\shorttitle{Dust Models for the Extinction toward SN 2014J}
\shortauthors{Gao et al.}

\begin{document}

\title{
%------------- enable for labelling preprint ---------------------------
%\vspace*{-2.0em}
%{\normalsize\rm Accepted for publication in
%               {\it Astrophys. J. Lett.}}\\
%\vspace*{1.0em}
Physical Dust Models for the Extinction toward Supernova 2014J in M82
\vspace*{1.0em}
}
%
%% Use \author, \affil, and the \and command to format
%% author and affiliation information.
%% Note that \email has replaced the old \authoremail command
%% from AASTeX v4.0. You can use \email to mark an email address
%% anywhere in the paper, not just in the front matter.
%% As in the title, use \\ to force line breaks.

\author{Jian Gao\altaffilmark{1,2},
        B.~W. Jiang\altaffilmark{1},
        Aigen Li\altaffilmark{2},
        Jun Li\altaffilmark{1}, and
        Xiaofeng Wang\altaffilmark{3}
        }
%\affil{Department of Astronomy, Beijing Normal University, Beijing
%100875, China; {\sf jiangao@bnu.edu.cn, bjiang@bnu.edu.cn}}
%\and
%\author{Aigen Li}
%\affil{Department of Physics and Astronomy, University of Missouri,
%Columbia, Missouri 65211, USA.} \email{lia@missouri.edu}

\altaffiltext{1}{Department of Astronomy,
                       Beijing Normal University,
                       Beijing 100875, China;
                       {\sf jiangao@bnu.edu.cn,
                            bjiang@bnu.edu.cn}
                      }
\altaffiltext{2}{Department of Physics and Astronomy,
                 University of Missouri, Columbia, MO 65211, USA;
                 {\sf lia@missouri.edu}
                }
\altaffiltext{3}{Department of Physics,
                 and Center for  Astrophysics,
                 Tsinghua University, Beijing 100084, China
                 {\sf wang$_{-}$xf@mail.tsinghua.edu.cn}
                 }
%%%%%%%%%%%%%%%%%%%%%%%%%%%%%%%%%%%%%%%

%
\begin{abstract}
Type Ia supernovae (SNe Ia)
are powerful cosmological ``standardizable candles''
and the most precise distance indicators.
However, a limiting factor in their use
for precision cosmology rests on our ability to
correct for the dust extinction toward them.
SN~2014J in the starburst galaxy M82,
the closest detected SN~Ia in three decades,
provides unparalleled opportunities
to study the dust extinction toward an SN Ia.
In order to derive the extinction
as a function of wavelength,
we model the color excesses toward SN 2014J, which are
observationally derived over a wide wavelength range
in terms of dust models consisting of a mixture
of silicate and graphite.
The resulting extinction laws steeply rise
toward the far ultraviolet, even steeper than
that of the Small Magellanic Cloud (SMC).
We infer
a visual extinction of $A_V \approx 1.9\magni$,
a reddening of $E(B-V)\approx1.1\magni$,
and a total-to-selective extinction ratio
of $R_V$\,$\approx$\,1.7,
consistent with that previously derived from
photometric, spectroscopic,
and polarimetric observations.
The size distributions of the dust
in the interstellar medium
toward SN 2014J are skewed
toward substantially smaller grains
than that of the Milky Way and the SMC.
\end{abstract}

%%%%%%%%%%%%%%%%%%%%%%%%%%%%
%%%%%%%%%%%%%%%%%%%%%%%%%%%%

\section{Introduction}
Type Ia supernovae (SNe Ia)
are considered to be one of the most
precise tools for determining
astronomical distances \citep{Howell11}.
Because of their high luminosity and relatively small
dispersion at the maxima of their bolometric light curves,
they are commonly utilized as
cosmological ``standardizable candles''.
The accelerated expansion of the Universe
and the presence of dark energy
were discovered through SNe Ia used as
standardizable candles
\citep{Riess98,Perlmutter99}.
The effectiveness of SNe Ia as distance indicators
and standard candles is hampered by
the systematic uncertainties related to
their explosion mechanism
and progenitor systems,
and more importantly,
the line-of-sight extinction. %\citep[see][]{Howell11}.
%
%This is because
The distance $d$
measured in parsec to a SN is
$\lg d = 0.2\left(m_\lambda-M_\lambda+5-A_\lambda\right)$,
where $m_\lambda$ and $M_\lambda$ are its apparent
and absolute magnitudes at wavelength $\lambda$,
and $A_\lambda$ is the extinction.
As it is not easy to directly measure $A_\lambda$,
one often measures the color excess (or reddening)
$E(\lambda-V)\equiv A_\lambda - A_V$,
where $A_V$ is the extinction
in the V-band (centered around $5500\Angstrom$).
SN reddening is often measured by comparing
the observed SN colors to
a zero-reddening locus. %\citep[e.g.,][]{Riess96,Phillips99}

\citeauthor{Cardelli89} (1989; CCM)
found that the Galactic extinction curves
(or extinction laws)
--- the wavelength dependencies
of the extinction --- can be closely parametrized
by the total-to-selective extinction ratio
$R_V\equiv A_V/E(B-V)$,
where the B-band centers around $4400\Angstrom$
\citep[also see][hereafter FTZ]{Fitzpatrick99}.
Astronomers often
derive $R_V$ for SNe Ia by fitting
%the photometric and/or spectroscopic data
the observed $E(\lambda-V)$
with the $R_V$-based CCM formula.
Once $R_V$ is determined,
one can apply the CCM-formula
(or some other parameterizations)
to derive $A_\lambda$.
However, we caution that the CCM-
and FTZ-parameterizations have been derived
for Galactic sightlines with $2 < \Rv < 5$,
and may not be valid for external galaxies.
Note that the CCM formula is not even
applicable to the Large and Small Magellanic Clouds
\citep[LMC, SMC;][]{Gordon03}.

SNe~Ia are so rare that nearby SNe~Ia ($d < 5\Mpc$)
are detected only about once a decade.
SN 2014J,
discovered %by \citet{Fossey14}
in the nearby starburst galaxy M82
at a distance of $d\approx 3.5\Mpc$
%$d_{\rm L}\approx 3.5\Mpc$
\citep{Dalcanton09},
is the nearest SN Ia
seen in the last three decades.
Its proximity offers an unprecedented opportunity
to study the extinction and reddening toward a SN Ia.
The aim of this Letter
is to derive $R_V$ and $A_\lambda$
%for SN 2014J %from the observed color-excess curve
%$E(\lambda-V)$ %without making a priori assumption
%of any template extinction law
%(\S\ref{sec:color_curve}).
%This is achieved
by fitting the reddening curve
%of SN 2014J
obtained by Amanullah et al.\ (2014)
during the epoch range of $[-5, +5]$ days
around its peak brightness
(\S\ref{sec:color_curve})
using the silicate-graphite model
(\S\ref{sec:model}).
The results are presented in \S3,
discussed in \S4, and summarized in \S5.

%%%%%%%%% Beginning Figure 1 %%%%%%%%%%%%%%%%%%%
\begin{figure}
\centering
\includegraphics[angle=0,width=3.6in]{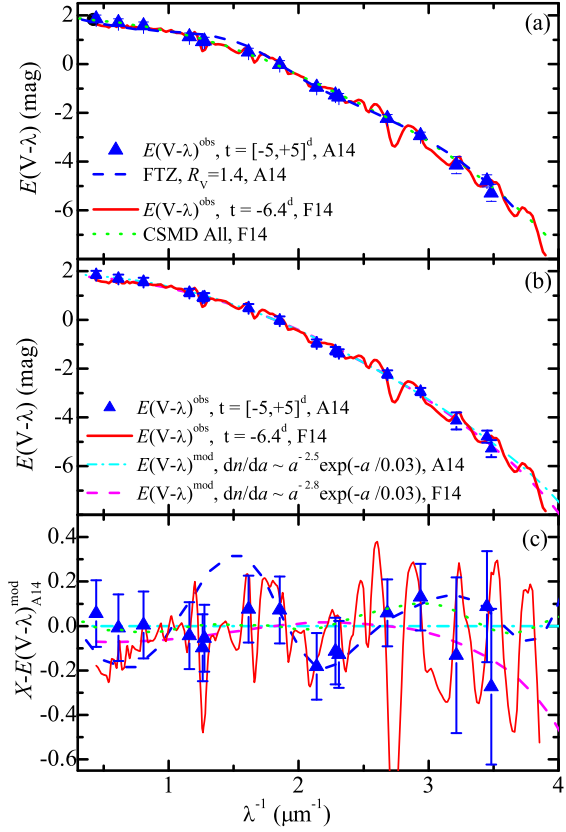}
\vspace{-1.0cm}
\caption{\footnotesize
        \label{fig:F14}
        (a) Comparison of the color-excess curve
            of F14 (red solid line)
            with that of A14 (blue triangles).
            The blue dashed line is the best-fit FTZ model of A14,
            while the green dotted line is the best-fit two-component
            model (i.e., a combination of interstellar reddening
            and circumstellar scattering)
            for all epochs (``CSMD All'') of SN 2014J of F14.
        (b) Comparison of the color-excess curves
            of F14 (red solid line) and A14 (blue triangles)
            with our model predictions
            (magenta dashed line for F14,
             and cyan dot-dashed line for A14).
        (c) Residuals between our model color-excess
            $E(V-\lambda)^{\rm mod}_{\rm A14}$ for A14 and
            (i) the data points of A14 (blue triangles),
            (ii) the FTZ fit of A14 (blue dashed line),
            (iii) the color-excess curve of F14 (red solid line),
            (iv) the CSMD fit of F14 (green dotted line), and
            (v) our model color-excess for F14 (magenta dashed line).
           }
\end{figure}
%%%%%%%%% Ending Figure 1 %%%%%%%%%%%%%%%%%%%

%%%%%%%%%%%%%%%%%%%%%%%%%%%%
\section{Color-Excess Curves of SN~2014J}\label{sec:color_curve}
Various studies have been carried out
to determine the $R_V$ value
for the sightline toward SN~2014J
\citep[e.g., see][]{Amanullah14,Foley14,Goobar14,Marion15,Welty14}.
More specifically,
based on the UV to near-IR photometry of SN 2014J
obtained with the WFC3 filters aboard
the {\it Hubble Space Telescope} (HST)
and ground-based telescopes,
%the {\it Nordic Optical Telescope}, and
%the {\it Mount Abu Infrared Telescope},
%
\citeauthor{Amanullah14} (2014, hereafter A14) determined
the reddening curve $E(\lambda-V)$
for SN 2014J over a wide wavelength range
by comparing the colors of SN~2014J
with that of SN~2011fe, a reddening-free SN Ia.
They derived $R_V\approx1.4\pm0.1$ by fitting
the observationally determined
$E(\lambda-V)$
with three template extinction laws:
an MW-type law as
parameterized by CCM or FTZ
(see Figure~\ref{fig:F14}a),
a third-order polynomial SALT2 law
\citep[see][]{Betoule14},
and a power-law parameterization
$A_\lambda\sim\lambda^{p}$
which was shown to be a good approximation
for multiple scattering scenarios \citep{Goobar08}.
%

%%%%%
We note that A14 shifted the effective wavelengths
of the HST/WFC3 filters (see their Figure~3).
In the following, we adopt the wavelength-shifted
color-excess data of A14.

Foley et al.\ (2014, hereafter F14)
determined the extinction curve toward SN~2014J
at $t = -6.4, +24.1\,d$ by comparing the UV/near-IR SED
of SN~2014J with that of SN~2011fe.
As shown in Figure~\ref{fig:F14},
the color-excess curve of F14
closely resembles that of A14.
The difference between the F14 curve
and that of A14, on average, is $\simali$0.19$\magni$
(see Figure~\ref{fig:F14}c).
Using FTZ and CCM models,
F14 derived $\Rv\approx1.66\pm0.03$
and $\approx1.44\pm0.03$
for SN 2014J, respectively.
F14 also argued that a two-component model (CSMD)
with both a circumstellar scattering component
of $A_\lambda\sim\lambda^{-2.57}$
and an FTZ interstellar reddening component
of %$E(B-V)\approx 0.45\magni$ and
$R_V\approx2.6$
could best account for the observed
properties of SN~2014J.

%%%%%%%%%%%%%%%%%%%%%%%%%%%%
\section{Dust Model}\label{sec:model}
We consider the silicate-graphite grain model
that consists of
%two separate components:
amorphous silicate and graphite
\citep{Draine84}.
We adopt the same exponentially cutoff
power-law size distribution for both components:
$dn_i/da = \nH B_i a^{-\alpha} \exp\left(-a/a_{b}\right)$
for the size range of
$50\Angstrom < a < 1\um$,
where $a$ is the spherical radius of the dust,
$\nH$ is the number density of H nuclei,
$dn_i$ is the number density of dust of type $i$
with radii in the interval [$a$, $a$\,$+$\,$da$],
$\alpha$ and $a_{b}$ are, respectively,
the power index and exponential cutoff size,
and $B_i$ is a constant related to
the total amount of dust of type $i$.
The total extinction at wavelength $\lambda$
is given by
\begin{equation}\label{eq:ext}
A_\lambda/\NH = \left(2.5\log e\right)
            \sum_i \int da\frac{1}{\nH}
            \frac{dn_i}{da}
            C_{{\rm ext},i}(a,\lambda) ~,
\end{equation}
where the summation is over the two grain types,
%(i.e., silicate and graphite),
%$\NH\equiv\int\nH\,dl$ is the H column density
%which is $\nH$
%%the H number density
%integrated over the line of sight $l$,
$\NH$ ($\nH$) is
the H column (number) density,
and $C_{{\rm ext},i}(a,\lambda)$
is the extinction cross section of
grain type $i$ of size $a$
at wavelength $\lambda$
calculated from Mie theory
using the optical constants of
\citet{Draine84}.

For a given set of parameters $\alpha$ and $a_{b}$,
we derive the constant $B_i$
from the abundances of the dust-forming elements.
Let $\xism$ be the total interstellar abundance
of element X (i.e., Fe, Mg, Si, O, and C)
relative to H in the interstellar medium (ISM) of M82,
$\xgas$ be the amount of X in the gas phase,
$\xdust$ be the amount of X contained in dust
(obviously, $\xdust$\,=\,$\xism-\xgas$),
and $\mux$ be the atomic weight of X.
Let $\rhosil\approx3.5\g\cm^{-3}$
and $\rhogra\approx2.24\g\cm^{-3}$
, respectively, be the mass density of
amorphous silicate and graphite.
For a chosen set of dust depletions,
we derive $B_i$ from
the dust size distributions:
\begin{equation}\label{eq:Csil}
\begin{split}
%\nH\Bsil = \frac{\mufe\fedust + \mumg\mgdust
%              + \musi\sidust + 4\times \mu_{\rm O}\sidust}
%        {\int da \left(4\pi/3\right) a^3\,\rhosil\,
%               a^{-\alpha}
%               \exp\left(-a/a_{b}\right)} ~~,
\nH\Bsil &= \frac{\mufe\fedust + \mumg\mgdust
              + \musi\sidust}
          {\int da \left(4\pi/3\right) a^3\,\rhosil\,
               a^{-\alpha}
               \exp\left(-a/a_{b}\right)}  \\
         & \quad \frac{4\times \mu_{\rm O}\sidust}{} ~~, \\
 \end{split}
\end{equation}
\begin{equation}\label{eq:cgra}
\nH\Bgra = \frac{\muc\cdust}
        {\int da \left(4\pi/3\right) a^3\,\rhogra\,
               a^{-\alpha}
               \exp\left(-a/a_{b}\right)} ~~,
\end{equation}
where we assume a stoichiometric composition of
Mg$_{\rm 2x}$Fe$_{\rm 2(1-x)}$SiO$_4$
for amorphous silicate.

M82 is a prototypical starburst galaxy,
experiencing a major star formation episode
in its nuclear region,
with strong superwind and SN activity.
\citet{Origlia04} obtained the stellar abundances
in the nuclear region of M82,
and compared them with those of the hot gas
derived from the nuclear X-ray spectra.
Compared with the solar abundance of \citet{Grevesse98},
both the cool stars and the hot gas in M82 suggest
a reduction of Fe/H by $\approx -0.35\pm0.2$ dex
(i.e., $\feism\approx 14.1\ppm$)
and an overall
${\rm Si/Fe}$ and ${\rm Mg/Fe}$ enhancement
by $\simali$0.4 and 0.5 dex, respectively
(i.e., $\siism\approx 35.5\ppm$,
$\mgism\approx 44.7\ppm$).
Oxygen is enhanced by $\simali$0.3 dex in stars
and reduced by $\simali$0.2 dex in the hot gas.\footnote{%
  An accurate knowledge of the O/H abundance is
  not required. The amount of O/H locked up
  in dust is controlled by Si/H:
  $\odust = 4\sidust$
  for a silicate composition of
  Mg$_{\rm 2x}$Fe$_{\rm 2(1-x)}$SiO$_4$.
  }
The stellar abundance of C derived by \citet{Origlia04}
is only $\simali$1/4 of solar
(i.e., $\cism\approx 83.2\ppm$).
%(i.e., a reduction of $\approx -0.60\pm0.10$ dex for C/H).

Similar to the Galactic ISM,
we assume in M82 that Fe, Mg and Si are
all locked up in silicate dust
(i.e., $\fedust\approx\feism\approx 14.1\ppm$,
$\mgdust\approx\mgism\approx 44.7\ppm$,
$\sidust\approx\siism\approx 35.5\ppm$;
\citealt{Origlia04}).
For carbon, it is less clear.
In the Galactic ISM,
a substantial fraction ($\approx 42\%$)\footnote{%
  We take the Galactic interstellar carbon abundance
  to be solar:
  $\cism\approx\csun\approx 331\ppm$
  \citep{Grevesse98}.
  }
of the total carbon abundance
is in the gas phase
($\cgas\approx140\ppm$, \citealt{Cardelli96}).
%
%
%
%In this work, for the carbon dust (i.e., graphite)
For carbon dust (i.e., graphite) in M82,
we will consider three cases
$\cdust = \cism \approx 83.2\ppm$
(i.e., all C is locked up in dust),
$\cdust = 1/2 \cism\approx 41.6\ppm$, and
%(i.e., 50\% of the C is locked up in dust), and
$\cdust = 0\ppm$
(i.e., all C is in the gas phase
and the dust model only consists of amorphous silicate).
With $\cdust = 41.6\ppm$, the model has
a silicate-to-graphite mass ratio of
$m_{\rm sil}/m_{\rm gra}\approx 10$.
This is close to that of the SMC
($m_{\rm sil}/m_{\rm gra}\approx 12$, Li et al.\ 2006).

To facilitate a direct comparison with
the color excesses $E(V-\lambda)^{\rm obs}$
derived by A14 for SN 2014J,
we first calculate $A_\lambda$ from Eq.\ref{eq:ext}
and then convert to reddening
$E(V-\lambda)^{\rm mod}\equiv A_V - A_\lambda$.
For simplicity, the model color-excess
has not been convolved with the HST/WFC3 filters.
We evaluate the goodness of fitting by
\begin{equation} \label{eq:2}
\frac{\chi ^2}{\rm d.o.f}
 = \frac{\sum {_{j=1}^{N_{\rm obs}}}
 \left[E(V-\lambda_j)^{\rm mod} - E(V-\lambda_j)^{\rm obs}\right]^{2}
 /\sigma(\lambda_j)^2}
 {N_{\rm obs} - N_{\rm para}} ~~,
\end{equation}
where $E(V-\lambda_j)^{\rm obs}$
is the observed color excess toward SN 2014J
at wavelength $\lambda_j$
derived by A14,
$\sigma(\lambda_j)$ is the uncertainty of
$E(V-\lambda_j)^{\rm obs}$,
$E(V-\lambda_j)^{\rm mod}$
is the model color excess at $\lambda_j$,
$N_{\rm obs} = 16$ is the number of observational data points,
and $N_{\rm para}$ is the number of adjustable parameters.
%

%%%%%%%%% Beginning Figure 2 %%%%%%%%%%%%%%%%%%%
\begin{figure}
%\epsscale{.50} \plotone{SAGE_CMD_JHK_SN1.ps}
\centering
\includegraphics[angle=0,width=3.2in]{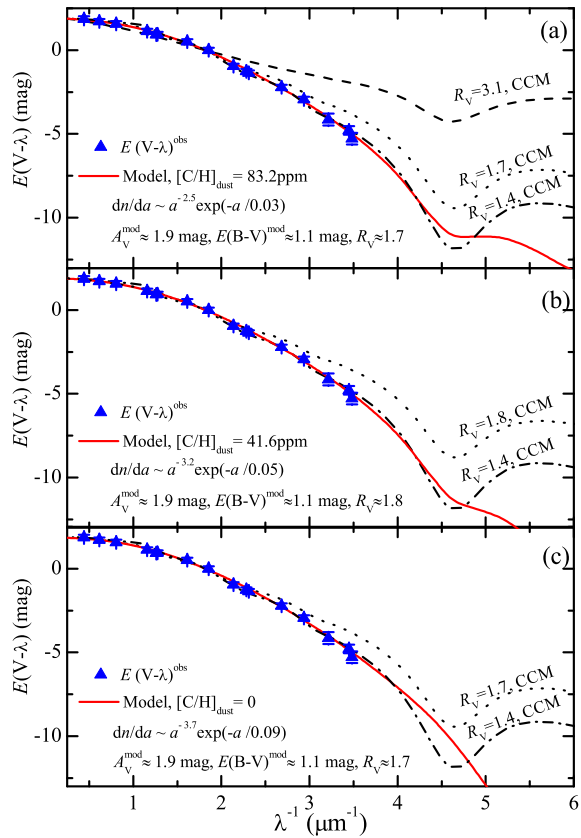}
\caption{\footnotesize
        \label{fig:A14}
         Fitting the color excesses
         between $-5$ and $+5$ days
         from the $B$ maximum of SN~2014J
         (blue triangles; A14)
         with the silicate-graphite model
         (red solid lines),
         assuming all (a), half (b), and none (c)
         of the carbon elements are depleted in graphite.
         The dotted lines plot the CCM reddening curves
         with the $R_V$ values corresponding to
         that of the models (a: $R_V$\,=\,1.7,
         b: $R_V$\,=\,1.8, c: $R_V$\,=\,1.7).
         For comparison, the Galactic average
         of $R_V$\,=\,3.1 (dashed line)
         and CCM reddening curve of $R_V$\,=\,1.4
         (dot-dashed line) are also shown.
         }
\end{figure}
%%%%%%%%% Ending Figure 1 %%%%%%%%%%%%%%%%%%%

%
\begin{table*}[htbp]
\renewcommand{\arraystretch}{1.5}
\vspace{-0.3cm}
\caption{Model Parameters and Results}
\vspace{-0.3cm}
\begin{center}
%\tiny
\begin{tabular}{cccccccccccc}
\hline\hline
$\cdust$\tablenotemark{a} & $\nH\Bgra$ & $\nH\Bsil$ &  $\alpha$ & $a_b$ & $\NH$ & $\chi^{2}/{\rm d.o.f}$ & $E(\rm B-\rm V)$ & $\Av$  & $\Rv$  \\
    (ppm)                 & ($\cm^{\alpha-1}/\HH$) & ($\cm^{\alpha-1}/\HH$)  &           &  ($\mum$) & ($10^{22}\cm^{-2}$) &  &($\magni$) & ($\magni$)  &  \\
\hline
 83.2                  & $1.3\times 10^{-19}$ & $4.0\times 10^{-20}$ & 2.5  & 0.03  & 1.3 & 0.45  & 1.1  & 1.9  & 1.7    \\
 41.6                  & $1.0\times 10^{-23}$ & $1.6\times 10^{-24}$ & 3.2  & 0.05  & 1.9 & 0.47  & 1.1  & 1.9  & 1.8    \\
 0                     & 0 & $1.2\times 10^{-26}$ & 3.7  & 0.09  & 3.3 & 0.53  & 1.1  & 1.9  & 1.7  \\
%\hline
 83.2\tablenotemark{b} & $3.7\times 10^{-23}$ & $1.4\times 10^{-23}$ & 3.1  & 0.05  & 1.6 & 0.51  & 1.1  & 2.0  & 1.8    \\
 41.6\tablenotemark{b} & $2.5\times 10^{-24}$ & $4.7\times 10^{-25}$ & 3.3  & 0.06  & 2.1 & 0.51  & 1.1  & 1.9  & 1.8    \\
\hline
 83.2\tablenotemark{c} & $4.7\times 10^{-19}$ & $1.4\times 10^{-19}$ & 2.4  & 0.03  & 1.3 & 0.43  & 1.1  & 1.9  & 1.7    \\
 83.2\tablenotemark{d} & $3.0\times 10^{-21}$ & $9.1\times 10^{-22}$ & 2.8  & 0.03  & 1.5 & 2.96  & 1.1  & 1.9  & 1.7    \\
\hline\hline
\end{tabular}
\vspace{-0.3cm}
\tablenotetext{1}{Assumed carbon depletion in graphite or amorphous carbon.}
\tablenotetext{2}{Amorphous carbon.}
\tablenotetext{3}{The F218W and F225W data points of A14 were excluded.}
\tablenotetext{4}{Best fit for F14 at $t=-6.4\,d$.}
\label{tab:results}
\end{center}
\end{table*}
%

%%%%%%%%% Beginning Figure 3 %%%%%%%%%%%%%%%%%%%
\begin{figure*}
%\epsscale{.50} \plotone{SAGE_CMD_JHK_SN1.ps}
\centering
\includegraphics[angle=0,width=5.6in]{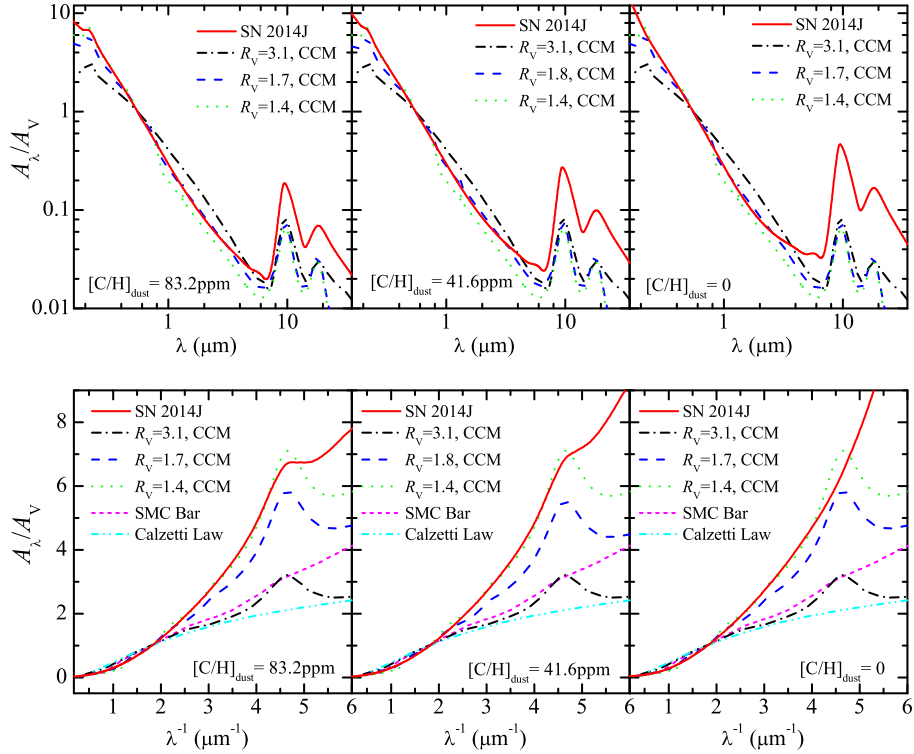}
\caption{\footnotesize
         \label{fig:extmod}
         Comparison of the model extinction curves
         of SN~2014J (red solid)
         with that for the CCM $R_V$\,=\,3.1
         (black dash-dotted), $R_V$\,=\,1.4 (green dotted),
         $R_V$\,=\,1.7 or 1.8 (blue dashed),
         the SMC bar (magenta short dashed),
         and the Calzetti attenuation law
         for starbursts (cyan dash-dot-dotted).
         }
\end{figure*}
%%%%%%%%% Ending Figure 3 %%%%%%%%%%%%%%%%%%%
%

%%%%%%%%% Beginning Figure 4 %%%%%%%%%%%%%%%%%%%
\begin{figure*}
%\epsscale{.50} \plotone{SAGE_CMD_JHK_SN1.ps}
\centering
\includegraphics[angle=0,width=5.8in]{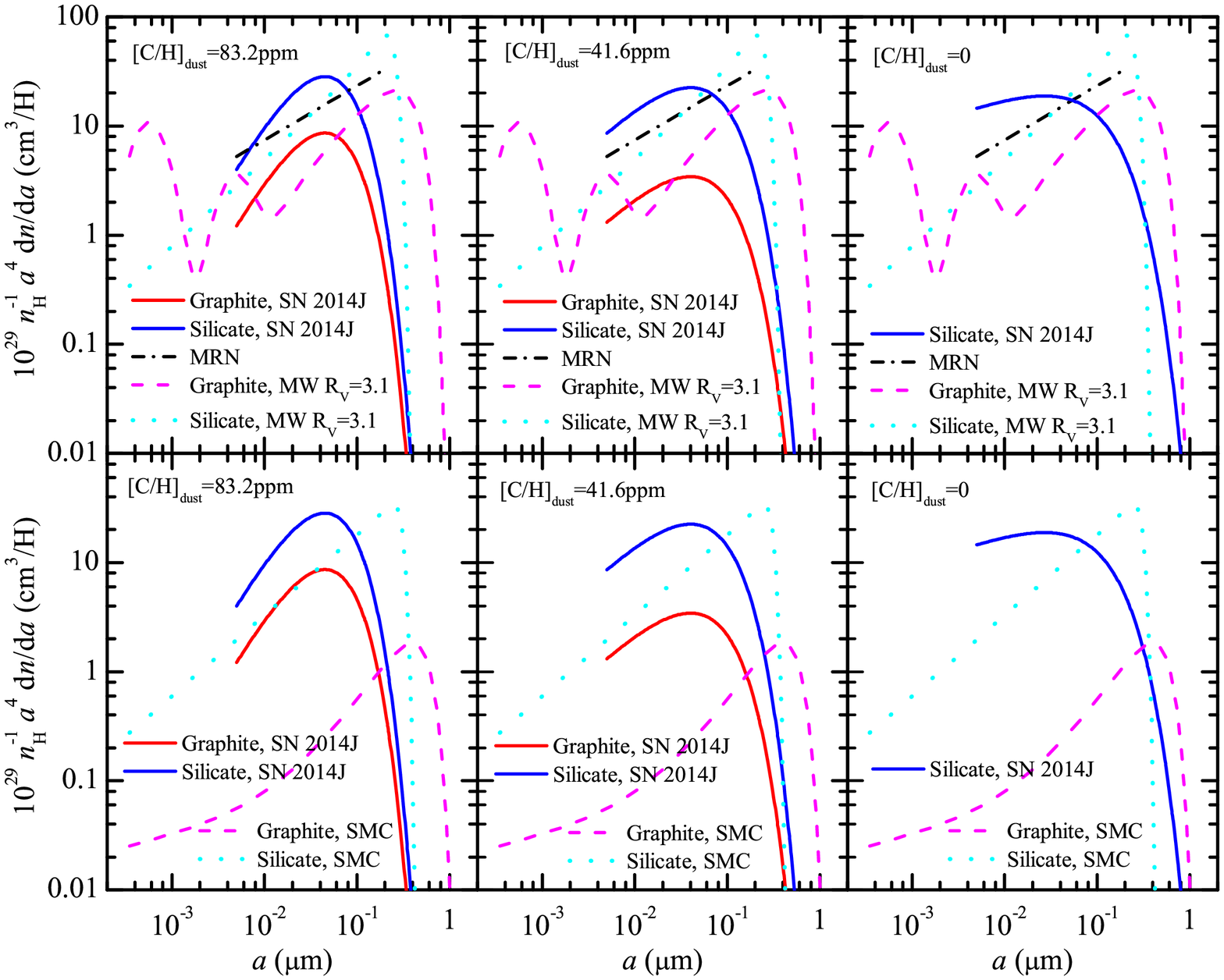}
\caption{\footnotesize
         \label{fig:dnda}
         Comparison of the size distributions
         of silicate (blue solid)
         and graphite (red solid)
         derived for SN 2014J
         %assuming different $\cdust$ values
         with that of the MW $R_V=3.1$ and SMC
         (silicate: cyan dotted;
          graphite: magenta dashed;
          see WD01).
         Also shown is the MRN size distribution
         (black dot-dashed).
         }
\end{figure*}
%%%%%%%%% Ending Figure 4 %%%%%%%%%%%%%%%%%%%

%%%%%% Sect. 3 %%%%%%
\section{Results}\label{sec:results}
In fitting the color excesses of SN 2014J
observationally determined by A14,
we have three parameters
(i.e., $N_{\rm para}=3$): $\alpha$, $a_b$, and $\NH$.
As shown in Figure~\ref{fig:A14},
excellent fits to the observed
color excesses can be achieved
by varying $\alpha$ and $a_b$
for a given $\cdust$
(see Table~\ref{tab:results}).

The best-fit models derive
$A_V \approx 1.9\magni$,
$E(B-V)\approx 1.1\magni$,
and $R_V$\,$\approx$\,1.7.
The reddening and $A_V$
are consistent with those reported earlier,
i.e., $E(B-V)$\,$\approx$\,0.8--1.3$\magni$
and $A_V$\,$\simali$1.8--2.0$\magni$
\citep{Amanullah14,Ashall14,Foley14,Goobar14,Welty14},
while the model $R_V$ values are somewhat larger
than most of the earlier results,
i.e., $R_V$\,$\simali$1.4
\citep{Amanullah14,Brown14,Goobar14,Marion15}.
The best-fit models suggest
$\NH$\,$\approx$\,1.3--3.3$\times10^{22}\cm^{-2}$
for the sightline toward SN~2014J,
somewhat higher than that of the hot gas in M82,
$\NH\approx\left(7.9\pm0.7\right)\times10^{21}\cm^{-2}$,
derived from {\it Chandra} observations
\citep{Origlia04}.

In Figure~\ref{fig:extmod},
we show the best-fit models
shown in Figure~\ref{fig:A14} %and Table~1
in terms of $A_{\lambda}/A_V$.
For comparison, we also show
the CCM reddening curve of $R_V$\,=\,1.4,
the average extinction curves
for the MW ($R_V=3.1$)
and the SMC bar. %\citep{Gordon03}.
The model extinction laws
for SN 2014J all exhibit a rapid far-UV rise
which is even much steeper than that of the SMC bar.
%On the other hand,
Unlike the SMC bar, the SN 2014J
extinction laws display an appreciable extinction bump
at 2175$\Angstrom$ for $\cdust>0$.
We also show in Figure~\ref{fig:extmod}
the extinction curves
predicted from the CCM formula
with the corresponding model-derived $R_V$ values
(i.e., $R_V$\,$\approx$\,1.7, 1.8).
It is seen that they
substantially differ from
that calculated from the dust models.

As mentioned in \S\ref{sec:color_curve},
A14 shifted the effective wavelengths
of the HST/WFC3 filters,
especially for the bluest F218W and F225W bands,
which are highly dependent
on the SN spectrum and the reddening law.
%(see their Figure~3).
%
To examine the effects of the wavelength-shifts,
we have also modeled the observed color-excess curve
of A14 by excluding the F218W and F225W data points
(which could be most affected).
The results do not show any appreciable differences
(see Table~\ref{tab:results}).
We have also modeled the F14 curve (at $t = -6.4\,d$).
%with the silicate-graphite model.
As shown in Figure~\ref{fig:F14}b
and Table~\ref{tab:results},
the model extinction curves and
the resulting $A_V$ and $R_V$
are very close to that derived from the A14 curve.
Figure~\ref{fig:F14}c shows the differences
between the best-fit model color-excesses
for A14 and that of the FTZ model for A14,
the CSMD model for F14,
as well as the best-fit model for F14.
They are generally within
the observational uncertainties.
However, as shown in Figure~\ref{fig:A14},
the model extinction at $\lambda < 0.3\mum$
differs considerably from the CCM parameterization.
The latter is known to be invalid for extragalactic
sightlines.

In Figure~\ref{fig:dnda}
we show the dust size distributions
derived from the models
with $\cdust$\,=\,83.2, 41.6 and 0$\ppm$.
Compared with that of the MW
average of $R_V=3.1$
[\citealt{Mathis77} (hereafter MRN);
\citealt{Weingartner01} (hereafter WD01)],
the size distributions of
the dust in the ISM toward SN 2014J are skewed
toward substantially smaller grains.
The MRN size distribution is a power-law
$dn/da\propto a^{-3.5}$
in the size range of
$50\Angstrom < a < 0.25\mum$
for both dust components.
The WD01 size distributions
extend the lower cutoff size
to $\amin=3.5\Angstrom$
with the smallest grains ($a\simlt50\Angstrom$)
constrained by the near- and mid-IR emission
(see \citealt{Li01}).
The dust model presented here for SN~2014J
assumes $\amin=50\Angstrom$
since the UV extinction cannot constrain
the exact size of nano-sized dust
(see \citealt{Li04}).

%%%%%%%%%%%%%%%%%%%%%%%%%%%%
\section{Discussion\label{sec:discussion}}

\subsection{The Extinction Curves\label{sec:extcurv}}

The host galaxy of SN 2014J, M82, is
regarded as one of the archetypical starburst
and superwind galaxies.
\citet{Calzetti94} derived the internal dust
extinction in starbursts
from their overall emission spectra.
The inferred attenuation curve
is much flatter than that derived for SN 2014J
(see Figure~\ref{fig:extmod}).
While the SN 2014J model extinction curves for
$\cdust>0$ exhibit an appreciable bump at 2175$\Angstrom$
(e.g., $\Delta\tau_{2175}\approx 1.8$
for the $\cdust\approx83.2\ppm$ model),
%see Table~\ref{tab:results}),
the starburst attenuation curve
shows no evidence for the 2175$\Angstrom$ bump.
It is not clear to what extent the flatness of
the apparent starburst attenuation curve
may be due to the effects of radiative transfer
in optically thick distributions of stars and dust.

\citet{Hutton14} analyzed the UV images
of M82 taken by the \emph{UV/Optical Telescope (UVOT)} on
board {\it Swift}.
The color-color diagram obtained
with the
UVW2 (2033$\Angstrom$),
UVM2 (2229$\Angstrom$), and
UVW1 (2591$\Angstrom$) filters
is especially sensitive to
the presence of
the 2175$\Angstrom$ bump.
%\citet{Hutton14}
They examined
the color-color diagram
and argued against a ``bump-less''
Calzetti-type law.
%instead, they were in favor of
%an extinction law
%with a 2175$\Angstrom$ bump.
%%

In the silicate-graphite model presented here,
the 2175$\Angstrom$ extinction bump is produced
by small graphite dust.
If we consider amorphous carbon \citep{Rouleau91}
instead of graphite,
the $2175\Angstrom$ bump will be absent.
Excellent fits can also be obtained from
a mixture of silicate and amorphous carbon
(see Table~\ref{tab:results}).
Unfortunately,
neither the A14 color-excess data points
nor the F14 extinction curve
covered the 2175$\Angstrom$ bump. Due to the lack of spectral features
(e.g., the 2175$\Angstrom$ bump),
we are not able to constrain the exact
composition and quantity
of the carbon dust component
(see Table~\ref{tab:results}).

The extinction laws derived for SN 2014J are
even steeper than that of the SMC bar
(see Figure~\ref{fig:extmod});
correspondingly, the dust sizes of SN~2014J
are smaller than that of the SMC bar
(see Figure~\ref{fig:dnda}).
This may be related to the intense UV radiation
and shocks associated with star formation in M82
that could destroy the dust
and lead to a predominance of small grains.

In addition to the 2175$\Angstrom$ bump,
the models presented here also predict two absorption
features around 9.7 and 18$\mum$ arising from
amorphous silicate (see Figure~\ref{fig:extmod}).
Depending on $\cdust$,
the optical depth of the 9.7$\mum$ feature
($\Delta\tau_{9.7}$)
ranges from $\simali$0.34 to $\simali$0.86.
\citet{Beirao08} reported the detection
of the 9.7$\mum$ feature
in the \emph{Spitzer}/IRS spectra
of the central region of M82,
with an optical depth of
$\Delta\tau_{9.7}$\,$\approx$\,0.3--3.1.
%\citet{Telesco15} obtained the $\simali$8--13$\mum$
%spectra of SN~2014J at 57, 81, 108, and 137 days
%after the explosion and did not detect
%the 9.7$\mum$ feature.
\citet{Telesco15} obtained
the $\simali$8--13$\mum$ mid-IR spectra
of SN 2014J at 57 to 137 days after explosion
and did not detect the 9.7$\mum$ feature.

%%%% R_V in Type Ia SNe %%%%
\subsection{%The Total-To-Selective Extinction Ratios:
            $R_V$\label{sec:RV}}
It is often suggested that the extinction
laws toward SNe Ia are ``non-standard'' or ``unusual''
in the sense that unlike the MW mean value
of $R_V\approx 3.1$,
SNe Ia often have a much smaller
$R_V$ (see Figure~3 of Howell 2011),
indicating steep UV extinction.

There are several examples of highly reddened
SNe~Ia for which $R_{V}$ can be measured directly.
They all have $R_{V} \lesssim 2$
\citep[e.g.,][]{Krisciunas06,Elias-Rosa06,
Elias-Rosa08,Nobili08,Wang08,Folatelli10}.
However, many low reddened
[$E(B-V ) < 0.3 \magni$] ones
have $R_V$ values close to that of
the MW \citep{Mandel11, Phillips12}.
On the other hand, statistical studies of
large samples of SNe Ia have found $R_{V} < 2$
\citep[e.g.,][]{Nobili05,Guy05,Hicken09,
Folatelli10,Burns11}.\footnote{%
  However, it has also been suggested that
  the low values of $R_{V}$ derived from
  large samples may partly result from
  the poor assumptions about the intrinsic
  color distribution of SNe Ia
  \citep[e.g., see][]{Foley11,Mandel11}.
  }
\citet{Betoule14} analyzed 740 low- and high-$z$
SNe Ia and yields $R_V \sim 2$.
%

%%%% R_V for SN 2014J %%%%

Many lines of evidence
show that the reddening law
to SN~2014J has a low value of $R_V\approx1.4$
(see \S\ref{sec:results}).
\citet{Patat14} and \citet{Kawabata14}
presented spectropolarimetric
and optical/near-IR multi-band polarimetric
observations of SN 2014J and both indicated
a low value of $\Rv<2$.
However, relatively larger $R_V$ values
have also been suggested for SN~2014J.
Using the equivalent widths of
10 diffuse interstellar bands,
\citet{Welty14} yielded
$E(B-V) \sim 0.71 \pm 0.11 \magni$
and $R_V\approx2.7$ for SN~2014J.

The extinction curves $A_\lambda/A_V$
and $R_V$ values derived here
for SN 2014J are based on detailed dust modeling
of the observed color excesses.
They are generally consistent with
previous studies of $R_V<2$ for SN 2014J.
However, we caution the use of $R_V$ to
derive an extinction law for SN~2014J
(or any extragalactic sightlines) since,
as demonstrated in Figure~\ref{fig:extmod},
the model extinction curves
differ substantially from that
calculated from the CCM formula.

\subsection{Interstellar or Circumstellar?\label{sec:iscs}}
It has been suggested that multiple scattering
by circumstellar dust surrounding their progenitors
could explain the non-standard reddening
observed in the lines of sight to SNe Ia
\citep{Wang05,Goobar08,Amanullah11}.
However, F14 argued that the wide range of
observed properties for SN~2014J may be caused by
a combination of interstellar reddening
and scattering off circumstellar material.
\citet{Johansson14}
analyzed the 3.6 and 4.5$\mum$
{\it Spitzer}/IRAC data of SN~2014J
and detected no significant IR excess.
They hence placed an upper limit of
$M_{\rm dust} \lesssim 10^{-5}\,M_{\odot}$
on the pre-existing dust
in the circumstellar environment of SN 2014J.
This is insufficient to account for
the observed non-standard reddening.
Moreover, \citet{Brown14} analyzed
the light curves and color evolution
obtained with {\it Swift}/UVOT.
They argued that these observations are
inconsistent with a contribution scattered
into the line of sight by circumstellar dust.
%
%

%%%%%%%%%%%%%%%%%%%%%%%%%%%%
\section{Conclusions}\label{conclusions}
The extinction toward SN 2014J in M82
is derived as a function of wavelength
from fitting the observed color excesses
with a mixture of silicate and graphite
or amorphous carbon dust.
Insensitive to the exact carbon dust
composition and quantity, the model
derives $A_V \approx 1.9\magni$,
$E(B-V)\approx1.1\magni$,
and $R_V$\,$\approx$\,1.7,
generally consistent with those
reported in the literature.

%%%%%%%%%%%%%%%%%%%%%%%%%%%%
\acknowledgments{
We thank the anonymous referee
and S.~Wang for helpful suggestions.
This work is supported by NSFC 11173007, 11373015, 11178003, and 11325313,
973 Program 2014CB845702 and 2013CB834903, NSF AST-1109039, NNX13AE63G,
and the Fundamental Research Funds for the Central Universities.
}

%%%%%%%%%%%%%%%%%%%%%%%%%%%%
%
\begin{table*}[htbp]
\renewcommand{\arraystretch}{1.5}
\vspace{-0.3cm}
\caption{Modeled Extinction towards SN 2014J\tablenotemark{a}}
\vspace{-0.3cm}
\begin{center}
%\tiny
\begin{tabular}{lccccccccccc}
\hline\hline
\multicolumn{2}{c}{}    & \multicolumn{2}{c}{$\cdust=41.6\rm ppm$}   &  \multicolumn{2}{c}{$\cdust=81.3\rm ppm$}             \\  \cline{3-6}
Band                    & $\lambda$   &$A_{\lambda}$  &   $C_{\rm ext}$     &$A_{\lambda}$ &   $C_{\rm ext}$           \\
                        & ($\mum$)    & ($\magni$)    &   (cm$^2/$H)    & ($\magni$)   &   (cm$^2/$H)          \\
\hline

Ly edge     &  0.091    &     39.87   &   2.06$\times10^{-21}$  &   29.32     &   2.22$\times10^{-21}$    \\
Ly$\alpha$ &  0.122    &     29.40   &   1.52$\times10^{-21}$  &   22.85     &   1.73$\times10^{-21}$    \\
UVW2/UVOT   &  0.203    &     13.87   &   7.16$\times10^{-22}$  &   13.09     &   9.92$\times10^{-22}$    \\
UVM2/UVOT   &  0.223    &     12.53   &   6.47$\times10^{-22}$  &   12.44     &   9.43$\times10^{-22}$    \\
UVW1/UVOT   &  0.259    &     8.64    &   4.46$\times10^{-22}$  &   8.63      &   6.54$\times10^{-22}$    \\
F225W/HST   &  0.287    &     6.98    &   3.60$\times10^{-22}$  &   6.96      &   5.28$\times10^{-22}$    \\
F275W/HST   &  0.290    &     6.84    &   3.53$\times10^{-22}$  &   6.82      &   5.17$\times10^{-22}$    \\
F218W/HST   &  0.311    &     5.97    &   3.08$\times10^{-22}$  &   5.96      &   4.52$\times10^{-22}$    \\
F336W/HST   &  0.340    &     5.03    &   2.59$\times10^{-22}$  &   5.02      &   3.81$\times10^{-22}$    \\
\emph{u}/SDSS      &  0.355    &     4.63    &   2.39$\times10^{-22}$  &   4.62      &   3.50$\times10^{-22}$    \\
U           &  0.365    &     4.39    &   2.27$\times10^{-22}$  &   4.39      &   3.33$\times10^{-22}$    \\
F438W/HST   &  0.433    &     3.15    &   1.63$\times10^{-22}$  &   3.16      &   2.40$\times10^{-22}$    \\
B           &  0.440    &     3.06    &   1.58$\times10^{-22}$  &   3.06      &   2.32$\times10^{-22}$    \\
F467M/HST   &  0.468    &     2.70    &   1.39$\times10^{-22}$  &   2.71      &   2.05$\times10^{-22}$    \\
\emph{g}/SDSS      &  0.469    &     2.69    &   1.39$\times10^{-22}$  &   2.70      &   2.05$\times10^{-22}$    \\
V           &  0.550    &     1.94    &   1.00$\times10^{-22}$  &   1.94      &   1.47$\times10^{-22}$    \\
F555W/HST   &  0.550    &     1.94    &   1.00$\times10^{-22}$  &   1.94      &   1.47$\times10^{-22}$    \\
\emph{r}/SDSS      &  0.617    &     1.54    &   7.93$\times10^{-23}$  &   1.53      &   1.16$\times10^{-22}$    \\
F631N/HST   &  0.630    &     1.48    &   7.62$\times10^{-23}$  &   1.47      &   1.12$\times10^{-22}$    \\
R           &  0.700    &     1.19    &   6.13$\times10^{-23}$  &   1.18      &   8.98$\times10^{-23}$    \\
\emph{i}/SDSS      &  0.748    &     1.04    &   5.35$\times10^{-23}$  &   1.03      &   7.82$\times10^{-23}$    \\
F814W/HST   &  0.792    &     0.92    &   4.75$\times10^{-23}$  &   0.92      &   6.95$\times10^{-23}$    \\
F845M/HST   &  0.863    &     0.77    &   3.99$\times10^{-23}$  &   0.77      &   5.83$\times10^{-23}$    \\
\emph{z}/SDSS      &  0.893    &     0.72    &   3.72$\times10^{-23}$  &   0.72      &   5.44$\times10^{-23}$    \\
I           &  0.900    &     0.71    &   3.66$\times10^{-23}$  &   0.71      &   5.35$\times10^{-23}$    \\
J/2MASS     &  1.235    &     0.38    &   1.94$\times10^{-23}$  &   0.37      &   2.82$\times10^{-23}$    \\
H/2MASS     &  1.662    &     0.22    &   1.14$\times10^{-23}$  &   0.22      &   1.64$\times10^{-23}$    \\
Ks/2MASS    &  2.159    &     0.15    &   7.53$\times10^{-24}$  &   0.14      &   1.07$\times10^{-23}$    \\
W1/WISE     &  3.353    &     0.08    &   4.32$\times10^{-24}$  &   0.08      &   5.83$\times10^{-24}$    \\
L           &  3.450    &     0.08    &   4.18$\times10^{-24}$  &   0.07      &   5.63$\times10^{-24}$    \\
$[3.6]$/IRAC&  3.545    &     0.08    &   4.06$\times10^{-24}$  &   0.07      &   5.45$\times10^{-24}$    \\
$[4.5]$/IRAC&  4.442    &     0.06    &   3.18$\times10^{-24}$  &   0.05      &   4.13$\times10^{-24}$    \\
W2/WISE     &  4.603    &     0.06    &   3.07$\times10^{-24}$  &   0.05      &   3.97$\times10^{-24}$    \\
M           &  4.800    &     0.06    &   2.96$\times10^{-24}$  &   0.05      &   3.81$\times10^{-24}$    \\
$[5.8]$/IRAC&  5.675    &     0.05    &   2.78$\times10^{-24}$  &   0.05      &   3.42$\times10^{-24}$    \\
$[8.0]$/IRAC&  7.760    &     0.10    &   5.20$\times10^{-24}$  &   0.07      &   5.60$\times10^{-24}$    \\
N           &  10.600   &     0.40    &   2.06$\times10^{-23}$  &   0.28      &   2.08$\times10^{-23}$    \\
W3/WISE     &  11.561   &     0.27    &   1.40$\times10^{-23}$  &   0.19      &   1.42$\times10^{-23}$    \\
Q           &  21.000   &     0.16    &   8.10$\times10^{-24}$  &   0.11      &   8.34$\times10^{-24}$    \\
W4/WISE     &  22.088   &     0.14    &   7.21$\times10^{-24}$  &   0.10      &   7.47$\times10^{-24}$    \\
\hline
\end{tabular}
\vspace{-0.3cm}
\tablenotetext{1}{The extinction results with $\cdust=41.6\rm ppm$ are recommended. Because of the space limitation of ApJL, this table is not shown in the published version of this Letter. } 
\label{tab:ext}
\end{center}
\end{table*}

\newpage
%%%%%%%%%%%%%%%%%%%%%%%%%%%%%%
%\renewcommand{\baselinestretch}{1.0}

%%%%%%%%%%%%%%%%%%%%%%%%%%%%

\end{document}